\documentclass[prb,twocolumn,showpacs,preprintnumbers,amsmath,amssymb,float]{revtex4-1}
\usepackage{graphicx}
\usepackage{amsmath}
\usepackage{epstopdf}
\usepackage{amsbsy}
\usepackage{pstricks}
\usepackage{color}
\usepackage{dcolumn}
\usepackage{bm}

\newcommand{\subfigimg}[3][,]{%
  \setbox1=\hbox{\includegraphics[#1]{#3}}
  \leavevmode\rlap{\usebox1}
  \rlap{\hspace*{0pt}\raisebox{\dimexpr\ht1-2\baselineskip}{#2}}
  \phantom{\usebox1}
  }

\begin{document}

\title{Electronic vortex structure of Fe-based superconductors: application to LiFeAs}

\author{B. Mencia Uranga$^1$, Maria N. Gastiasoro$^1$, and Brian M. Andersen$^1$}
\affiliation{$^1$Niels Bohr Institute, University of Copenhagen, Universitetsparken 5, DK-2100 Copenhagen,
Denmark
}

\date{\today}

\begin{abstract}

Detailed tunneling spectroscopy of vortex core states can provide important insight to the momentum structure of the superconducting order parameter. We present a theoretical study of vortex bound states in iron-based superconductors by use of a realistic five-band model relevant to these systems, and superconductivity stabilized by spin-fluctuation generated pairing vertices yielding an $s\pm$ gap structure. The computed local density of states agrees remarkably well with both the bias dependence of the local conductance and the spatial structure of the low-bias conductance as obtained by scanning tunneling microscopy measurements on LiFeAs [T. Hanaguri {\it et al.}, Phys. Rev. B {\bf 85} 214505 (2012)].

\end{abstract}

\pacs{74.20.-z, 74.25.Uv, 74.55.+v, 74.70.Xa}

\maketitle

\section{introduction}

The understanding of the microscopic origin of electron pairing in strongly correlated electron systems remains an ultimate goal in the field of unconventional superconductivity. Generating Cooper pairs from purely repulsive bare Coulomb interactions, as for example in terms of exchange of antiferromagnetic spin excitations, requires distinctive properties of the resulting gap symmetry as given, for example, by a sign-changing superconducting gap in momentum space. Thus, a detailed experimental determination of the superconducting gap structure and its evolution with e.g. temperature and electron filling is of crucial importance in order to unveil the constituents that drive the superconducting instability. 

Local perturbations of the superconducting condensate as found, for example, near defect sites or vortex cores, constitute an important means to study the gap symmetry. This is because the low-energy states in the vicinity of such regions are highly dependent on the properties of the gap structure of the bulk superconducting phase.\cite{balatsky,fischer} In the case of Fe-based superconductors, scanning tunneling spectroscopy (STS) studies of bound, or quasi-bound, states near impurity sites have mapped out a rich series of tunneling spectra that still awaits quantitative theoretical modelling.\cite{hoffman11,Song,grothe12,yang,chi14,yin15} Complicating factors in this endeavor include the multi-band electronic structure of these materials and the orbital degrees of freedom of the scattering centers, resulting in substantial parameter dependence of the modelling.\cite{nakamura,gastiasoroprb13,gastiasoro13} The scattering potential from vortex cores, on the other hand, does not contain orbital complexity/uncertainty, since it is entirely set by the quantum flux lines generated by the external field. In this respect, it may be more straightforward to deduce properties of the host superconducting gap symmetry from the measured tunneling conductance near vortex cores. 

In Fe-based superconductors a limited number of experimental studies have measured the detailed sub-gap tunneling conductance near vortex cores.\cite{yin,shan,Song,hanaguri12} For example Hanaguri {\it et al.}\cite{hanaguri12} detected and mapped out the voltage and real-space dependence of the sub-gap vortex core states in LiFeAs.\cite{hanaguri12} This material is particularly suitable for surface-sensitive probes due to its  atomically flat nonpolar cleaved surfaces. In addition LiFeAs is superconducting in its stoichiometric composition, i.e. without the complications of dopant disorder, and it exhibits neither a magnetic nor a structural transition. In the bulk homogeneous phase, LiFeAs exhibits a typical two-gap density of states spectrum with the inner (outer) gap edge at 2.9 meV (6.0 meV)\cite{hanaguri12} composed, presumably, of mainly $d_{xy}$ ($d_{xz}$/$d_{yz}$) orbital character. The latter follows from a comparison of typical DFT bandstructure results and the momentum-resolved gap structure as measured by ARPES. \cite{borisenko12,Hajiri,Umezawa} 

The main results of the STS study near the vortex cores of LiFeAs in Ref.~\onlinecite{hanaguri12} include a discrete set of resonant in-gap states with a pronounced conductance peak around -0.9 meV, i.e. just below the Fermi level $E_F$, and a smaller and broader conductance peak near -2.3 meV [see Fig.~\ref{fig:3}(a) and Ref.~\onlinecite{hanaguri12}]. The peaks exhibit symmetry partners at positive bias, but significant anisotropy exists in their spectral weight. The conductance peak at -0.9 meV disperses away from $E_F$ with increasing distance from the vortex core center and smoothly approaches the inner superconducting gap. The spatial profile of the low-energy core states were found to consist of a four-fold symmetric star shape with high LDOS tails along the nearest As directions, i.e. the 110 directions of the 1-Fe unit cell. At biases beyond the lowest energy peak (at -0.9 meV) the conductance tails split into parallel streaks and become less pronounced [see Fig.~\ref{fig:4} and Ref.~\onlinecite{hanaguri12}].

Here we study the electronic properties with particular focus on the low-energy sub-gap states in vortex cores in realistic models relevant to Fe-based superconductors. We include all five $d$ orbitals of the Fe sites, and stabilize superconductivity from spin-fluctuation exchange resulting in an $s\pm$-wave pairing state. The main findings of this approach is a remarkable resemblance of the theoretically obtained vortex core spectrum to that measured by Hanaguri {\it et al.}\cite{hanaguri12} Based on this agreement and the absence of any tuning parameters of the scattering potential, we conclude that the applied bandstructure and superconducting pairing gap provide a good description of the superconducting phase of LiFeAs.  

Several earlier theoretical vortex state studies of Fe-based superconductors exist in the literature. Some have focussed on the effect of competing magnetism and pairing symmetry dependence of the vortex core electronic structure within simplified two-orbital models.\cite{Hu,Jiang,Araujo,Gao,mishra} Other works have performed a more systematic study of the vortex core bound state spectrum with the number of bands.\cite{DaWang} In a recent publication, we have utilized a five-band model including multi-orbital Hubbard correlations to investigate vortex-induced stripe magnetism which allowed us to explain recent observed field-enhanced magnetism in Ba(Fe$_{0.95}$Co$_{0.05}$)$_2$As$_2$ as seen by neutron scattering and $\mu$SR experiments.\cite{larsen14} Wang {\it et al.}\cite{Wang} studied the LDOS anisotropy of core states in LiFeAs within a quasi-classical one-band approach, and concluded that the four-fold star shape observed by Hanaguri {\it et al.}~\cite{hanaguri12} is a Fermi surface anisotropy effect, and not caused by gap anisotropy. Our calculations presented below within a five-band self-consistent approach support this interpretation of the data in LiFeAs. Finally it should be noted that the experiment by Song {\it et al.}~\cite{Song} in which they found highly anisotropic vortices in FeSe, has stimulated several theoretical studies of vortices in this material.\cite{Chowdury,Hung,QEWang}

Before entering the model and result sections, we note that a more quantitative understanding of the vortex cores spectrum in LiFeAs is particularly desirable given the substantial discussion of potential alternative pairing structures of this material.\cite{pjh15} Specifically, the less nested band of this compound\cite{Borisenko10} and the fact that the largest measured gap appears on the smallest hole pocket around the $Z$ point, has led to a controversy about the pairing origin, and the distribution of signs on the various Fermi pockets.\cite{platt,kreisel,kotliar,ahn13,saito1,saito2,nourafkan} Thus it is currently of interest to include other experimental probes to help resolve the detailed gap structure of LiFeAs in particular, and Fe-based superconductors in general. 

\section{model}

\begin{figure}[b]
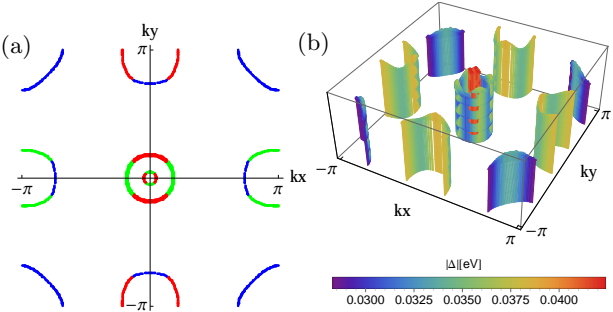

\centering
\subfigimg[width=0.22\textwidth]{\raisebox{10pt}{ \hspace*{-10pt} (a)}}{FS_LiFeAs}
\subfigimg[width=0.22\textwidth]{\raisebox{2.5pt}{\hspace*{-10pt} (b)}}{Gap_LiFeAs}
\caption{(a) Fermi surface with the dominant orbital character indicated by the color notation green $(d_{yz})$, red $(d_{xz})$, and blue $(d_{xy})$. (b) Absolute value of the superconducting gap at the Fermi surface.}
\label{fig:1}
\end{figure}

The starting point for the theoretical modelling is the following five-orbital Hamiltonian
\begin{equation}
 \label{eq:H}
 H=H_{0}+H_{BCS}+H_{Z},
\end{equation}
where $H_0$ contains the kinetic energy given by a tight-binding fit to the DFT bandstructure,\cite{gastiasoroprb13} including hopping integrals of all Fe orbitals to fifth nearest neighbors
\begin{equation}
 \label{eq:H0}
H_{0}=\sum_{\mathbf{ij},\mu\nu,\sigma}t_{\mathbf{ij}}^{\mu\nu}e^{i \varphi_{\mathbf{ij}} }c_{\mathbf{i}\mu\sigma}^{\dagger}c_{\mathbf{j}\nu\sigma}-\mu_0\sum_{\mathbf{i}\mu\sigma}n_{\mathbf{i}\mu.\sigma}.
\end{equation}
Here, the operators $c_{\mathbf{i} \mu\sigma}^{\dagger}$ create electrons at the $i$-th site in orbital $\mu$ and spin $\sigma$, and $\mu_0$ is the chemical potential used to set the doping $\delta=\langle n \rangle - 6.0$.
The indices $\mu$ and $\nu$ run from 1 to 5 corresponding to the Fe orbitals $d_{3z^2-r^2}$, $d_{yz}$, $d_{xz}$, $d_{xy}$, and $d_{x^2-y^2}$, respectively. The corresponding Fermi surface is shown in Fig.~\ref{fig:1}(a) in the one Fe unfolded Brillouin zone. It consists of three hole pockets (two smaller $\Gamma$-centered circular hole pockets and one $M$-centered larger and more squarish hole pocket), and two electron pockets at $X$ and $Y$.  

The presence of an external magnetic field is described by standard means by use of the Peierls phases
$\varphi_{\mathbf{ij}}=\frac{-\pi}{\Phi_{0}}\oint_{\mathbf j}^{\mathbf i} \mathbf{A} \cdot  \mathbf{dr}$, 
where $\Phi_{0}=\frac{h}{2e}$ is the half flux quantum and the integral is a line integral along the straight line joining lattice sites
$\mathbf{j}$ and $\mathbf{i}$.

The second term in Eq.~(\ref{eq:H}) is given by
\begin{equation}
 H_{BCS}=-\sum_{\mathbf{i}\neq \mathbf{j},\mu\nu}[\Delta_{\mathbf{ij}}^{\mu\nu}c_{\mathbf{i}\mu\uparrow}^{\dagger}c_{\mathbf{j}\nu\downarrow}^{\dagger}+\mbox{H.c.}],
\end{equation}
with superconducting order parameter defined by $\Delta_{\mathbf{ij}}^{\mu\nu}=\sum_{\alpha\beta}\Gamma_{\mu\alpha}^{\beta\nu}(\mathbf{r_{ij}})\langle\hat{c}_{\mathbf{j}\beta\downarrow}\hat{c}_{\mathbf{i}\alpha\uparrow}\rangle$. Here $\Gamma_{\mu\alpha}^{\beta\nu}(\mathbf{r_{ij}})$ denotes the effective pairing strength between sites (orbitals) $\mathbf{i}$ and $\mathbf{j}$ ($\mu$, $\nu$, $\alpha$ and $\beta$) obtained from the RPA spin- $\chi^{RPA}_s$ and charge susceptibilities $\chi^{RPA}_c$ relevant for LiFeAs \cite{gastiasoro13}
\begin{align}
\Gamma_{\mu\alpha}^{\beta\nu}({\mathbf{k}}&-{\mathbf{k}}')=\left[ \frac{3}{2} U^s \chi^{RPA}_s({\mathbf{k}}-{\mathbf{k}}') U^s +  \frac{1}{2} U^s\right.\nonumber\\
&\left.-\frac{1}{2} U^c \chi^{RPA}_c({\mathbf{k}}-{\mathbf{k}}') U^c +  \frac{1}{2} U^c \right]_{\mu\alpha}^{\beta\nu},
\end{align}
where $U^s$ and $U^c$ are $5\times5$ matrices identical to those of Ref.~\onlinecite{gastiasoroprb13}. 
The real-space pairings are then obtained by $\Gamma_{\mu\alpha}^{\beta\nu}(\mathbf{r_{ij}})=\sum_{\mathbf{q}} \Gamma_{\mu\alpha}^{\beta\nu}({\mathbf{q}}) \exp(i{\mathbf{q}}\cdot({\mathbf{r_i}}-{\mathbf{r_j}}))$ where we retain
all possible orbital combinations up to next-nearest neighbors (NNN). For the present band, the RPA susceptibilities are strongly peaked near $(0,\pm\pi)$ and $(\pm\pi,0)$ favoring an $s^\pm$ pairing state. A recent theoretical spin fluctuation study of pairing in LiFeAs used a full 3D ARPES-derived bandstructure, and also found standard $s_{\pm}$-wave pairing to be the dominant instability.\cite{kreisel} In the current case, the resulting gap structure is shown in Fig.~\ref{fig:1}(b) where we plot the amplitude of the gap on the various Fermi surface sheets. As seen, the largest gap exists on the smallest inner hole pocket in agreement with experiments. Below, for numerical finite-size reasons we have to use a gap that is larger than the experimental case in order to enhance the spectral resolution at low energies. This, however, causes only minor quantitative changes in the obtained LDOS (for example it makes the inner gap less pronounced).

The last term in Eq.~(\ref{eq:H}), $H_Z$, is the Zeeman term accounting for spin-dependent energy shifts due to the external magnetic field
\begin{equation}
\label{Hz}
H_{Z}= h \sum_{\mathbf{i} \mu} (n_{\mathbf{i} \mu \uparrow}-n_{\mathbf{i} \mu \downarrow}).
\end{equation}
Here, $ h=-\frac{\mu_{B}g_{s}B}{2}$, $\mu_B$ is the Bohr magneton and $g_s$ is electron g-factor.

Performing a standard Bogoliubov transformation applied to Eq.~(\ref{eq:H}) leads to the following multi-band Bogoliubov de-Gennes (BdG) equations\cite{gastiasoro13}
\begin{align}
\sum_{\mathbf{j}\nu}
\begin{pmatrix}
H^{\mu\nu}_{\mathbf{i} \mathbf{j} \sigma} & \Delta^{\mu\nu}_{\mathbf{i} \mathbf{j}}\\
\Delta^{\mu\nu*}_{\mathbf{i} \mathbf{j}} & -H^{\mu\nu*}_{\mathbf{i} \mathbf{j} \bar{\sigma}}
\end{pmatrix}
\begin{pmatrix}
 u_{\mathbf{j}\nu}^{n} \\ v_{\mathbf{j}\nu}^{n}
\end{pmatrix}=E_{n}
\begin{pmatrix}
 u_{\mathbf{i}\mu}^{n} \\ v_{\mathbf{i}\mu}^{n}
\end{pmatrix},
\end{align}
where
\begin{align}
 H^{\mu\nu}_{\mathbf{i} \mathbf{j} \sigma}&=t_{\mathbf{ij}}^{\mu\nu} e^{i \varphi_\mathbf{ij}}+\delta_{\mathbf{ij}}\delta_{\mu\nu}[-\mu_0+ \sigma h].
 \end{align}
We find the stable solutions through iterations of the following self-consistency equations 
\begin{eqnarray}
\langle n_{\mathbf{i}\mu\uparrow} \rangle&=&\sum_{n}|u_{\mathbf{i}\mu}^{n}|^{2}f(E_{n}),\\ 
\langle n_{\mathbf{i}\mu\downarrow} \rangle\!&=&\!\sum_{n}|v_{\mathbf{i}\mu}^{n}|^{2}(1\!-\!f(E_{n})),\\
\Delta_{\mathbf{ij}}^{\mu\nu}&=&\sum_{\alpha\beta}\Gamma_{\mu\alpha}^{\beta\nu}(\mathbf{r_{ij}})\sum_{n}u_{\mathbf{i}\alpha}^{n}v_{\mathbf{j}\beta}^{n*}f(E_{n}), 
\end{eqnarray}
where $\sum_n$ denotes summation over all eigenstates $n$. The self-consistency is unrestricted and the fields are allowed to vary on each site and orbital.

Below, the lattice constant $a$ is chosen as the unit of length, and we apply the Landau gauge $\mathbf{A}(\mathbf{r})=(By,0)$, which corresponds to a magnetic field $\mathbf{B}=B (-\mathbf{\hat e_{z}})$. The magnetic translation operators (MTO), which commute with the Hamiltonian, are given as follows \cite{Goto}
\begin{equation}
\label{MTO}
 \mathcal{M}_{\mathbf R} \hat \psi(\mathbf r)=e^{-i \frac{1}{2} \chi(\mathbf r,\mathbf R ) \sigma_{z}} \hat \psi(\mathbf r - \mathbf R ),
\end{equation}
where $\sigma_{z}$ is the Pauli matrix, $\hat \psi(\mathbf r)$ are the wave functions of the quasiparticles with $u$ and $v$ components, $\chi(\mathbf r,\mathbf R)=\frac{2 \pi}{\Phi_{0}} \mathbf A (\mathbf R) \cdot \mathbf r$ and $\mathbf R= m N_{x} \mathbf{\hat e_{x}}+n N_{y} \mathbf{\hat e_{y}}$ with $m,n$ integers and $N_{x}$,$N_{y}$ being the dimensions of the magnetic unit cell (MUC). 
In order to have MTOs that fulfill the composition law $\mathcal{M}_{\mathbf R_{m}} \mathcal{M}_{\mathbf R_{n}}= \mathcal{M}_{\mathbf R_{m} + \mathbf R_{n}}$, it is required that the MUC contains an even number of half flux quantum $\Phi_{0}$. The magnetic field is fixed such that the flux going through the MUC is $\Phi=2 \Phi_{0}$. The fulfilment of the composition law leads to the generalized Bloch theorem, which reads
\begin{equation}
\label{Bloch}
 \mathcal{M}_{\mathbf R} \hat{\psi}_{\mathbf{k}}(\mathbf r)= e^{-i \mathbf k \cdot \mathbf R} \hat{\psi}_{\mathbf{k}}(\mathbf r),
\end{equation}
where $\mathbf k $=$\frac{2 \pi l_{x}}{N_{x}} \mathbf{\hat e_{x}}+\frac{2 \pi l_{y}}{N_{y}} \mathbf{\hat e_{y}}$ with $l_{x,y}=0,1,...,N_{x,y}-1$ are the wave vectors defined in the first Brillouin zone of the vortex lattice and $ \hat{\psi}_{k}(\mathbf r)$ denote eigenstates of the Hamiltonian and the MTO. By use of Eq.~(\ref{MTO}) and Eq.~(\ref{Bloch}), the eigenfunctions of the Hamiltonian transform under translations as
\begin{align}
\begin{pmatrix}
 u_{\mathbf{i+R}\mu}^{n} \\ v_{\mathbf{i+R}\mu}^{n}
\end{pmatrix}
=e^{i \mathbf k \cdot \mathbf R}
\begin{pmatrix}
e^{-i \frac{1}{2}\chi(\mathbf r,\mathbf R )} u_{\mathbf{i}\mu}^{n} \\ e^{i \frac{1}{2} \chi(\mathbf r,\mathbf R )} v_{\mathbf{i}\mu}^{n}
\end{pmatrix},
\end{align}
where $\mathbf i$ takes values in the magnetic unit cell and $\mathbf r$=$\mathbf{i+R}$.
 Note that since a minimum of two superconducting flux quanta need to penetrate the MUC, the magnetic field is related to the real-space system size by $B \sim \frac{58500}{N_{x}N_{y}}T$ ($a=2.66 \text{\AA}$ for LiFeAs). 
 For the five-band model used here, we are restricted numerically to systems of sizes less than $(N_{x},N_{y})=(56,28)$, which we use in the present calculation, indicating that we have a field of $\sim37 T$, whereas a $0.5T$ field was used in the experiments of Ref.~\onlinecite{hanaguri12}. Despite this substantial quantitative discrepancy, we can still reliably study the properties of a single vortex since the neighboring vortices cause only minor quantitative effects. However, in order to make better contact to the experimental conditions by Hanaguri {\it et al.}, we set $B=0.5T$ in $H_z$ of Eq.~(\ref{Hz}) in order to avoid unphysically large Zeeman energy splittings. This is only a quantitative issue and does not influence the main points of the theoretical results discussed in the following sections.

\begin{figure}[tb]
\centering
\subfigimg[width=0.45\textwidth]{
\raisebox{0pt}{ \hspace*{-10pt} (a)}
\raisebox{-65pt}{ \hspace*{-21pt} (b)}
\raisebox{-115pt}{ \hspace*{-21pt} (c)}
\raisebox{-170pt}{ \hspace*{-21pt} (d)}
\raisebox{-230pt}{ \hspace*{-21pt} (e)}
\raisebox{-231pt}{\hspace*{17pt}  y}
\raisebox{-292pt}{\hspace*{48pt}  x}}{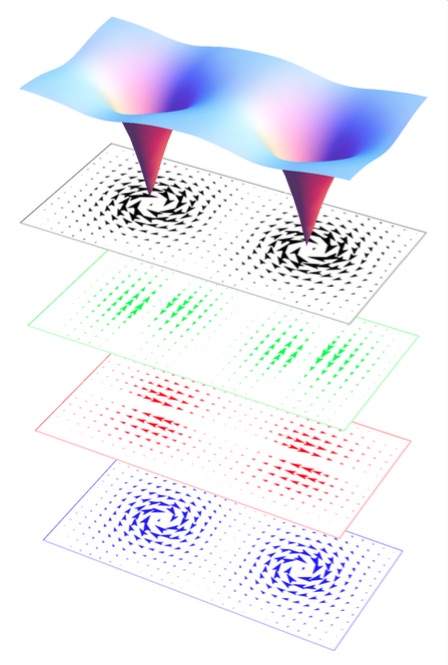}

\caption{(a) Real-space plot of the amplitude of the superconducting order parameter $|\Delta(\mathbf{r_i})|$ for magnetic flux $\Phi=2\Phi_{0}$ through the unit cell. The cores exhibit the usual suppression of $|\Delta(\mathbf{r_i})|$. (b-e) Real-space structure of the total (black) and orbitally resolved (green, red, and blue) supercurrents. The currents $d_{yz}$ (green), $d_{xz}$ (red) and $d_{xy}$ (blue) have the arrowheads amplified for visual clarity.}
\label{fig:2}
\end{figure}

Figure~\ref{fig:2}(a) shows the self-consistent superconducting order parameter for $\Phi=2 \Phi_{0}$.
The vortex cores generated by the external field can be clearly seen by the two suppressed regions of $\Delta(\mathbf{r_i})$. Here, $\Delta(\mathbf{r_i})$ refers to the superconducting order parameter at each site defined by
\begin{equation}
\Delta(\mathbf{r_i}) \equiv \frac{1}{9}\sum_{\mu \nu,\mathbf{j*}} \Delta_{\mathbf{ij*}}^{\mu\nu}e^{i \varphi_{\mathbf{ij*}}},
\end{equation}
where the index $\mathbf{j*}$ includes the set of onsite, nearest neighbor and next-nearest neighbor lattice sites to site $\mathbf{i}$.

Figure~\ref{fig:2}(b-e) show the total and orbitally resolved supercurrents obtained from the expression \footnote{This expression is obtained form the continuity equation $\frac{\partial \rho_{i}}{\partial t}+ (\mathbf{\nabla} \cdot \mathbf{j})_{i}=0$, where $\rho_{i}=-e n_{i}$ and  $\mathbf{j}$ are the electron density and current operators, respectively. The time derivative of the density operator at site $i$ and orbital $\mu$ is given by $\dot{n}_{i \mu}= \frac{i}{\hbar} [H, n_{i \mu}]$. The Hamiltonian entering the commutator to the full BCS Hamiltonian (i.e. prior to the mean-field decoupling). One is left with $\dot{n}_{i \mu}= \frac{i}{\hbar} [H_{0}, n_{i \mu}]$, whose thermal average leads to Eq.~(\ref{current}).}

\begin{align}
\label{current}
\langle (\mathbf{\nabla} \cdot \hat{\mathbf{j}})_{i \mu} \rangle &= \nonumber \\
-\frac{ie}{\hbar}
& \sum_{j \nu n}  [t_{ji}^{\nu \mu}e^{i \varphi_{ji}}
(  u_{j \nu }^{n*} u_{i \mu }^{n} f(E_{n}) + v_{j \nu }^{n} v_{i \mu }^{n*} f(-E_{n})  ) \nonumber \\
&-t_{ij}^{\mu \nu}e^{i \varphi_{ij}}
(  u_{i \mu }^{n*} u_{j \nu }^{n} f(E_{n}) + v_{i \mu }^{n} v_{j \nu }^{n*} f(-E_{n})  )].
\end{align}

The main contribution to the supercurrents comes from intra-orbital terms of the orbitals that dominate the Fermi surface, i.e. the $d_{yz}$, $d_{xz}$, and $d_{xy}$ orbitals, and hence only these contributions are shown in Fig.~\ref{fig:2}(c-e). The currents from the other two $e_g$ orbitals are negligible. 

\begin{figure}[]
\subfigimg[width=0.223\textwidth]{\raisebox{20pt}{ \hspace*{-10pt} (a)}}{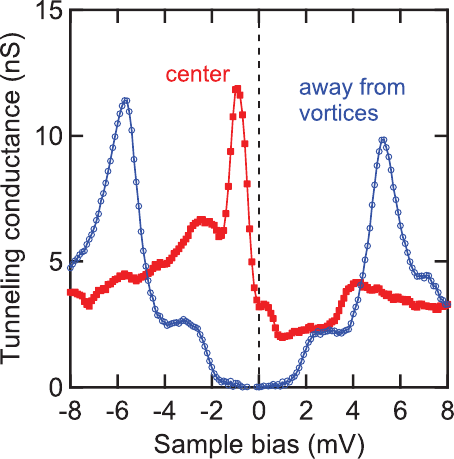}
\hspace*{-5pt}\raisebox{-2.5pt}{\subfigimg[width=0.255\textwidth]{\raisebox{20pt}{\hspace*{0pt} (b)}}{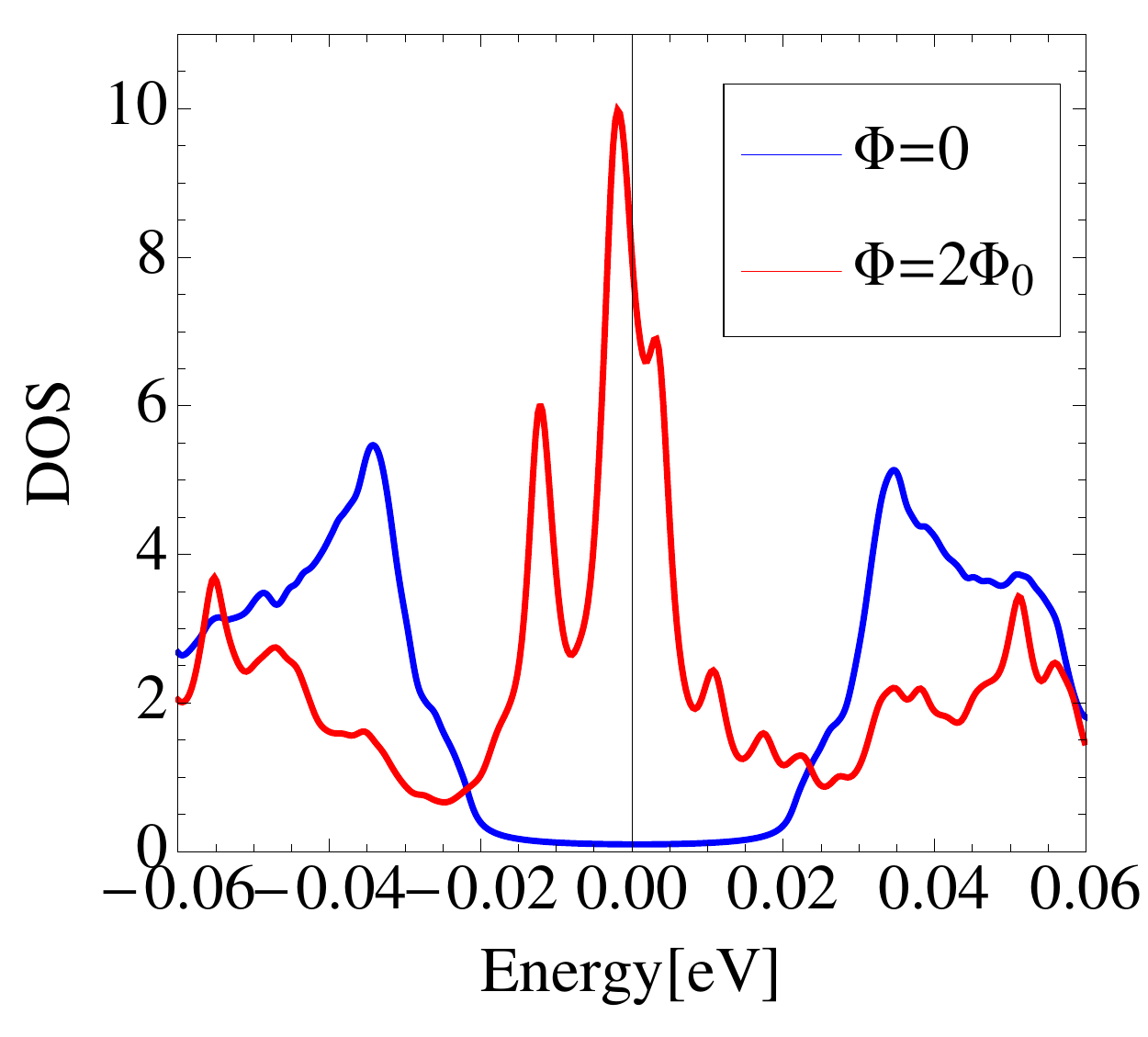}}
\hspace*{-4pt}\subfigimg[width=0.255\textwidth]{\raisebox{25pt}{\hspace*{0pt} (c)}}{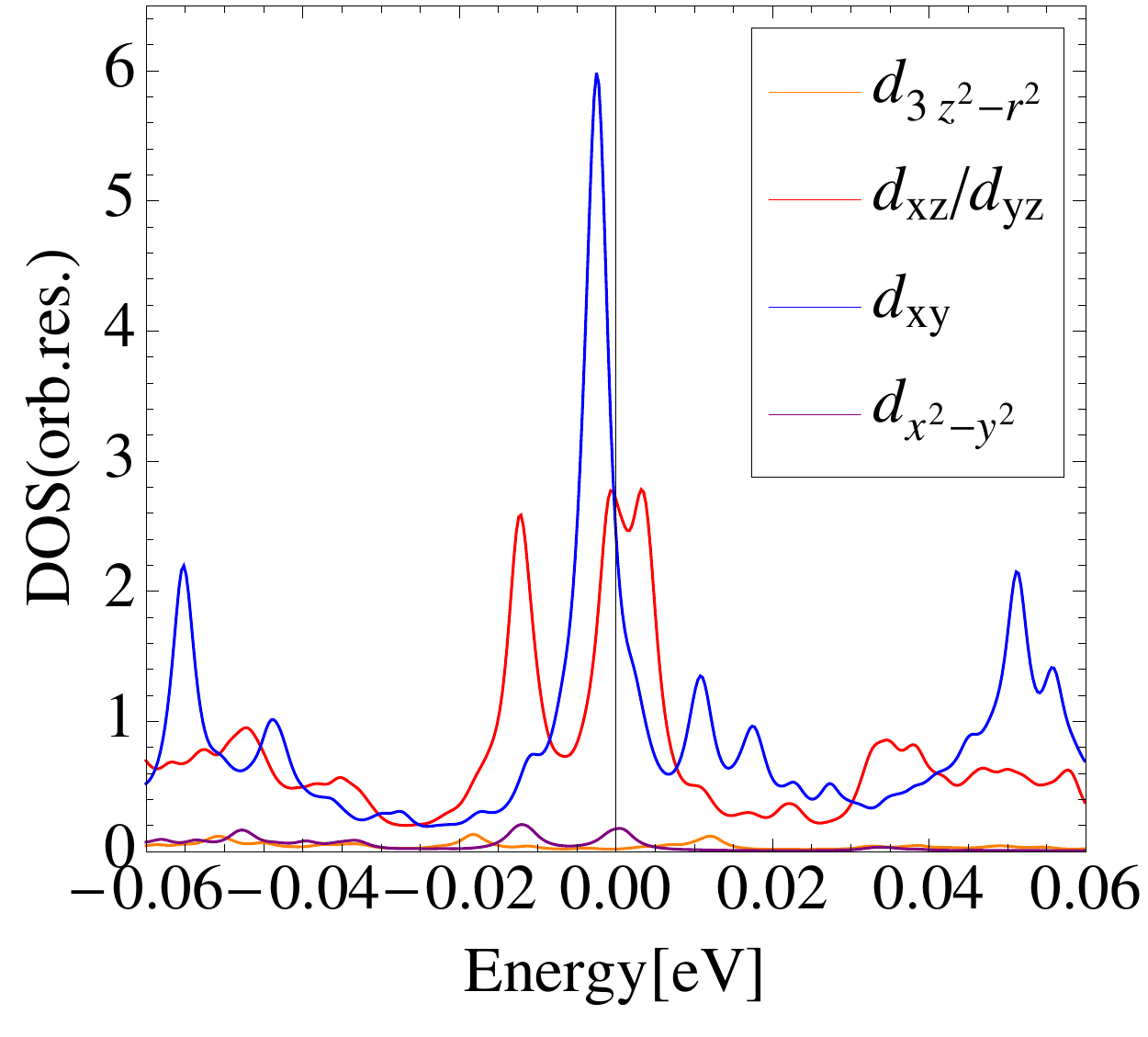}
\hspace*{-3pt}\raisebox{18.7pt}{\subfigimg[width=0.2\textwidth]{
\raisebox{22pt}{\hspace*{-15pt} (d)}
\raisebox{-94pt}{\hspace*{23pt} \fontsize{7pt}{7pt}\selectfont Energy[eV]}
\raisebox{-85.18pt}{\hspace*{-75pt} \fontsize{7pt}{7pt}\selectfont -0.06}
\raisebox{-85pt}{\hspace*{60pt} \fontsize{7pt}{7pt}\selectfont 0.06}
\raisebox{-85pt}{\hspace*{-57pt} \fontsize{7pt}{7pt}\selectfont 0}
\raisebox{-30pt}{\hspace*{-63pt}  x}}{vortexline}}
\raisebox{19.5pt}{\hspace*{-5pt}\subfigimg[width=0.035\textwidth]{}{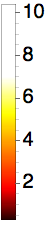}}
\caption{(a) Experimental tunneling spectra from Hanaguri {\it et al.}\cite{hanaguri12} in LiFeAs taken at the center of the vortex core (red) and away from vortex core region (blue). (b) Total LDOS calculated at the center of the vortex (red) and away from vortices (blue) to be compared to panel (a). (c) Orbitally resolved DOS calculated at the center of the vortex corresponding to the red curve in panel (b). (d) Total LDOS along a line cut of 28 lattice sites through the vortex core.}
\label{fig:3}
\end{figure}

Figure~\ref{fig:3}(a) shows the measured conductance in LiFeAs at the vortex center (red) and away from vortices (blue), reproduced from the publication of Hanaguri {\it et al.}\cite{hanaguri12}. In Fig.~\ref{fig:3}(b) we show the calculated local density of states (LDOS) $\rho(i,\omega)$ given by
\begin{equation}
\rho(i,\omega)=-\frac{1}{\pi} \mbox{Im}  \sum_{n \mu} \left[ 
\frac{\vert u_{i \mu }^{n} \vert^{2}}{w-E_{n}+i \eta} +
\frac{\vert v_{i \mu }^{n} \vert^{2}}{w+E_{n}+i \eta}
\right] ,
\end{equation}
at the vortex center (red) and away from vortices (blue) obtained within the current model. As seen from comparison to Fig.~\ref{fig:3}(a), we find good agreement between theory and experiment; in both cases there are two clear sub-gap conductance peaks at negative biases with the most pronounced peak just below the Fermi level. These two conductance peaks have particle-hole symmetric partners at positive biases which are, however, strongly suppressed by their associated coherence factors. 

To extract information about the orbital content of the sub-gap peaks, we plot the orbitally resolved calculated LDOS in Fig.~\ref{fig:3}(c). Evidently the main inner peak just below the Fermi level consists of mainly $d_{xy}$ character whereas the outer peak consists of $d_{xz}$ and $d_{yz}$ orbital states. We note that as opposed to the case of impurity bound states, there is no low-energy contribution to the LDOS from the $e_g$ orbitals since the vortex states are low-energy Andreev-like states whereas the impurity bound states can be generated from high-energy states.\cite{gastiasoroprb13,gastiasoro13} Finally, in Fig.~\ref{fig:3}(d) we show the spatial evolution of the total LDOS along a line cut through the cortex core. The main in-gap bound state is seen by the bright spot in the center of the plot. When moving away from the core center, the bound state disperses to larger energies and merges with the inner gap edge similar to the experimental finding.~\cite{hanaguri12} 

\begin{figure}[t]
\centering
\subfigimg[width=0.4\textwidth]{
\raisebox{-20pt}{ \hspace*{0pt} (a)}
\raisebox{-20pt}{ \hspace*{85pt} (b)}
\raisebox{-222pt}{ \hspace*{-143pt} y}
\raisebox{-250pt}{ \hspace*{16pt} x}
\raisebox{-222pt}{ \hspace*{63pt} y}
\raisebox{-250pt}{ \hspace*{14pt} x}
}{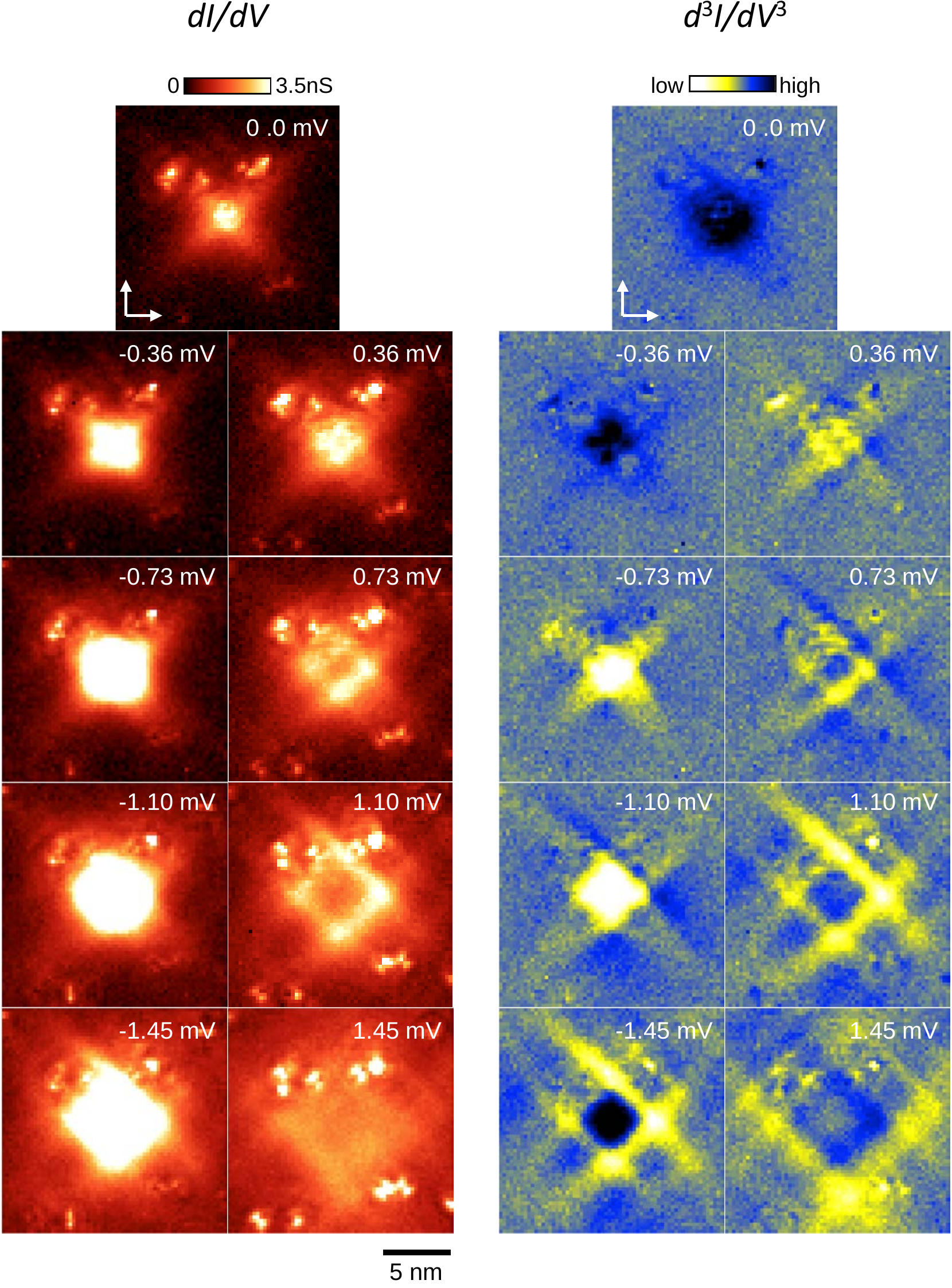}
\caption{(a) Measured spatial dependence of the tunneling conductance probing the inner sub-gap peak at both negative and positive biases obtained by Hanaguri {\it et al.}\cite{hanaguri12} (b) In the second derivative of the conductance one is able to identify a splitting of the tails of the star, and the splitting appears to set in earlier (in bias voltage) for positive biases versus negative biases (compare e.g. the panels at $\pm 0.73$ mV.}
\label{fig:4}
\end{figure}

We turn now to a discussion of the spatial profile of the conductance at fixed low biases. Figure~\ref{fig:4}(a) shows the results of the tunneling conductance measurements around a single vortex at low biases inside the fully gapped region.\cite{hanaguri12} As seen, the conductance displays a four-fold star shape with weight leaking out along the 110 directions. Away from the Fermi level, each of the four tails split up into two sub tails as seen most clearly from Fig.~\ref{fig:4}(b). This tail-splitting appears to set in earlier (in bias) for positive bias than for negative bias as evident for example from comparison of the panels at $\pm 0.73$ mV or $\pm 1.10$ mV in Fig.~\ref{fig:4}(b).

\begin{figure}[t]
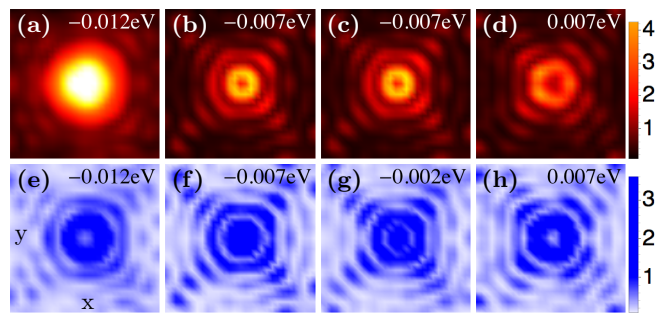


 \renewcommand{\arraystretch}{0}

 \begin{tabular}{@{}c@{}c@{}c@{}c@{}c}
\hspace*{-4pt} 
\subfigimg[width=0.115\textwidth]{\raisebox{-10pt}{ \hspace*{-3pt} \color{white}{\textbf{(a)}}}}{ld_m12_10}&
\subfigimg[width=0.115\textwidth]{\raisebox{-10pt}{ \hspace*{-3pt} \color{white}{\textbf{(b)}}}}{ld_m7_10}&
\subfigimg[width=0.115\textwidth]{\raisebox{-10pt}{ \hspace*{-3pt} \color{white}{\textbf{(c)}}}}{ld_m7_10}&
\subfigimg[width=0.115\textwidth]{\raisebox{-10pt}{ \hspace*{-3pt} \color{white}{\textbf{(d)}}}}{ld_7_10}&
\subfigimg[width=0.02\textwidth]{\raisebox{-10pt}{ \hspace*{-3pt} }}{scale_tot}\\
\hspace*{-4pt}  
\subfigimg[width=0.115\textwidth]{
\raisebox{-10pt}{ \hspace*{-6pt} \textbf{(e)}}
\raisebox{-30pt}{ \hspace*{-23pt} y}
\raisebox{-56pt}{ \hspace*{11pt} x}}{ld_m12_orb_10}&
 \subfigimg[width=0.115\textwidth]{\raisebox{-10pt}{ \hspace*{-3pt} \textbf{(f)}}}{ld_m7_orb_10}&
\subfigimg[width=0.115\textwidth]{\raisebox{-10pt}{ \hspace*{-3pt} \textbf{(g)}}}{ld_m2_orb_10}&
 \subfigimg[width=0.115\textwidth]{\raisebox{-10pt}{ \hspace*{-3pt} \textbf{(h)}}}{ld_7_orb_10}&
 \subfigimg[width=0.02\textwidth]{\raisebox{-10pt}{ \hspace*{-3pt} }}{scale_orb}
  \end{tabular}

\caption{(a-d) Spatial dependence ($28\times 28$ sites) of the total LDOS calculated around a single vortex core at selected representative energies inside the gap. (e-h) Same as (a-d) but only the $d_{xy}$ contribution to the LDOS is shown. The star-shape of the LDOS is a property of the $d_{xy}$ orbital, and is evident in the total LDOS only when this orbital contribution dominates.}
\label{fig:5}
\end{figure}

Figure~\ref{fig:5}(a-d) shows a representative set of calculated sub-gap total LDOS patterns around a vortex core. Similar to the experimental results in Fig.~\ref{fig:4}(a,b), the LDOS exhibit a four-fold star shape with tails along the 110 directions which eventually disappears as the energy becomes comparable to the inner gap edge. With increasing energy, we also find LDOS features that resemble that each tail splits up into two sub tails whereas the more squarish LDOS pattern found experimentally does not seem to be reproduced theoretically. The tail-splitting takes place initially at positive energy as seen by comparing the panels at energies $\pm 0.007$ eV. 

What is the origin of the star-shaped low-energy LDOS? The spatial profile of core states are known to unveil nodes in the gap, for example, as seen in the case of cuprates.\cite{fischer,udby2006} That family of materials, however, is known to be prone to competing order which complicates the understanding of the core states due to locally induced spin- and charge density waves.\cite{arovas,andersen2000,chen,zhu02,ghosal2002,takigawa,udby2006,Schmid10,andersen11} The gap structure displayed in Fig.~\ref{fig:1}(b) exhibits no (accidental) gap nodes but rather a Fermi surface anisotropy with mimima located mainly along the nearest-neighbor Fe-Fe 100 directions. This is certainly true for the $d_{xy}$-dominated large hole pocket near the $M$ point, which is also the main orbital character of the inner gap edge in the homogeneous case, and the lowest most pronounced peak in the vortex cores. Thus, the gap structure at the $d_{xy}$-dominated Fermi surface does not explain the star shape. In Fig.~\ref{fig:5}(e-h) we show the spatially resolved $d_{xy}$-contribution to the LDOS, revealing that the star shape is a property of the $d_{xy}$ orbital states [As discussed above, the $d_{xz}$ and $d_{yz}$ do not contribute much at this energy, and we have additionally verified that they do not exhibit "star quality"]. From the modelling one may also conclude that the spatially resolved total LDOS exhibits the four-fold star shape only at the energies at which $d_{xy}$ dominates the LDOS. In addition to gap anisotropy, Wang {\it et al.}\cite{Wang} recently studied the role of Fermi surface-anisotropy on the core states. When applied to LiFeAs, they concluded that the square shape of the large hole pocket near  $M=(\pi,\pi)$ with significant flat regions along the 100 directions [see also Fig.~\ref{fig:1}] feed into the spatial structure of the core states, and cause a four-fold star shaped pattern similar to experiment.

\begin{figure}[t]

\hspace*{-5pt}\subfigimg[width=0.223\textwidth]{\raisebox{20pt}{ \hspace*{-3pt} (a)}}{FS_Ikeda}
\hspace*{-5pt}\raisebox{-11pt}{\subfigimg[width=0.255\textwidth]{\raisebox{26.5pt}{\hspace*{0pt} (b)}}{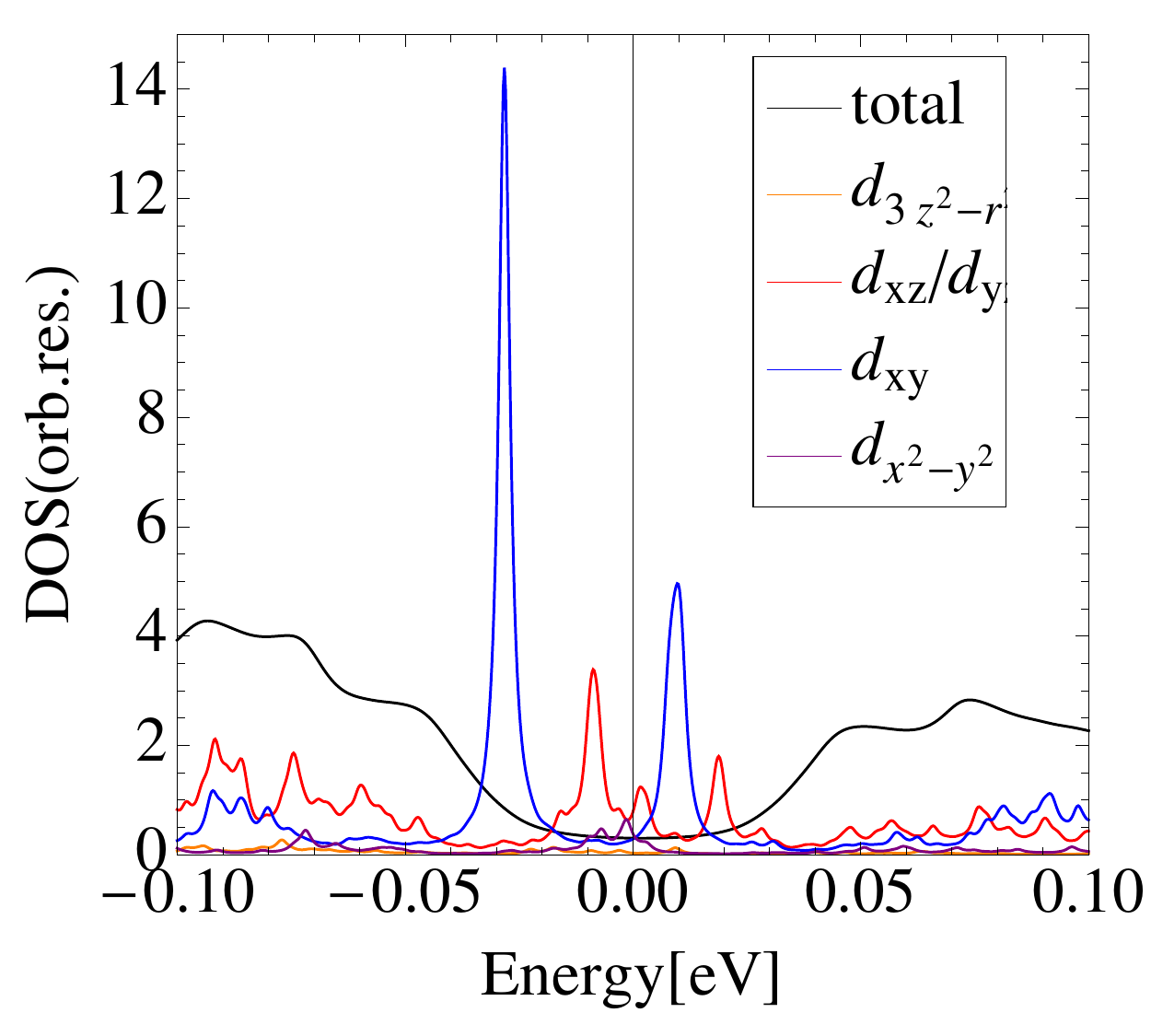}}
\hspace*{-0pt}\raisebox{20pt}{\subfigimg[width=0.2\textwidth]{
\raisebox{25pt}{\hspace*{0pt} (c)}
\raisebox{-31pt}{ \hspace*{-25pt} \color{white}{y}}
\raisebox{-78pt}{ \hspace*{34pt} \color{white}{x}}}{ld_m28_Ikeda_10}}
\hspace*{-2pt}\raisebox{18pt}{\subfigimg[width=0.035\textwidth]{}{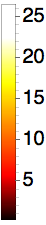}}
\hspace*{-0pt}\raisebox{20pt}{\subfigimg[width=0.2\textwidth]{
\raisebox{24pt}{\hspace*{-15pt} (d)}
\raisebox{-31pt}{ \hspace*{-10pt} \textbf{y}}
\raisebox{-78pt}{ \hspace*{33.5pt} \textbf{x}}}{ld_m28_orb_Ikeda_10}}
\hspace*{-2pt}\raisebox{15.5pt}{\subfigimg[width=0.035\textwidth]{}{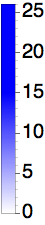}}

\caption{(a) Fermi surface of a band with a circular $(\pi,\pi)$ pocket,\cite{ikeda} showing the orbital majority with the same color code as in Fig.~\ref{fig:1}(a). (b) Orbitally resolved LDOS calculated at the center of the vortex core. (c) Spatially resolved total LDOS ($28\times 28$ sites) at the pronounced sub-gap energy $-0.028$ eV. (d) Same as (c) but only displaying the $d_{xy}$ contribution to the LDOS.}
\label{fig:6}
\end{figure}

We have verified that the origin of the star-shaped LDOS obtained in our model is also caused by Fermi surface anisotropy, rather than gap anisotropy. Specifically, it is precisely the squarish form of the $d_{xy}$-dominated hole pocket as seen from Fig.~\ref{fig:1}(b) that gives rise to the 110 tails. In Fig.~\ref{fig:6}(a) we show the Fermi surface obtained by Ikeda {\it et al.}\cite{ikeda}, displaying largely the same topology and orbital content as the one used above with the exception of a roughly circular $d_{xy}$ hole pocket at $M$. From the LDOS shown in Fig.~\ref{fig:6}(b) one sees directly a significant band dependence of the sub-gap vortex bound states, but in the present case the orbital polarization of the bound states is even more pronounced. Zooming in on the $d_{xy}$-dominated bound state at -0.028 eV, we find a perfect rotationally symmetric total {\it and} $d_{xy}$-resolved LDOS as seen from Fig.~\ref{fig:6}(c) and Fig.~\ref{fig:6}(d), respectively. Thus, in this case the different directions of the Fermi velocities of the $M$-centered hole pocket are equally weighted and hence the LDOS star shape has vanished. 

\section{Conclusions}

In summary we have performed a fully self-consistent real-space BdG study of vortex core states in five-orbital models relevant to Fe-based superconductors. The superconducting order was stabilized by spin fluctuation-derived pairing vertices generating an $s_{\pm}$-wave gap structure. By application to LiFeAs we find striking agreement with STS measurements by Hanaguri {\it et al.} on this compound.\cite{hanaguri12} In particular the details of the energy dependence and the spatial structure seems in almost quantitative agreement without any tuning parameters. From this fact we conclude that our model provides a reasonable description of LiFeAs, without the necessity to invoke more exotic paring states of this material. In future theoretical studies it would be interesting to compute the vortex core bound states within other pairing states for LiFeAs and compare their vortex core spectrum to experiments. 

\section{Acknowledgements}

We acknowledge useful discussions with T. Hanaguri, P. J. Hirschfeld, and A. Kreisel. The authors thank T. Hanaguri for providing us with the original data set used in
Ref.~\onlinecite{hanaguri12}. B. M. U. acknowledges support from La Caixa scholarship. B. M. A. and M. N. G. were supported by the Lundbeckfond fellowship (grant A9318).

\end{document}